\documentclass[useAMS,usenatbib]{mn2e}
\usepackage{graphicx}                                            
\usepackage{times}
\usepackage{epsfig}
\usepackage{rotating}
\usepackage{longtable}

\def\kms{\mbox{${\rm km}\:{\rm s}^{-1}\:$}}

\def\lesssim{\mathrel{\hbox{\rlap{\hbox{\lower4pt\hbox{$\sim$}}}\hbox{$<$}}}}
\let\la=\lesssim
\def\gtrsim{\mathrel{\hbox{\rlap{\hbox{\lower4pt\hbox{$\sim$}}}\hbox{$>$}}}}
\let\ga=\gtrsim

\title[The black hole candidate MAXI J1659-152 ]{The black hole candidate MAXI J1659-152: spectral and timing analysis during its 2010 outburst}
\author[T. Mu\~noz-Darias et al.]{T.~Mu\~noz-Darias$^{1}$, S.~Motta$^{1,2}$, H.~Stiele$^{1}$ and T.~M.~Belloni$^{1}$\\
$^{1}$INAF-Osservatorio Astronomico di Brera, Via E. Bianchi 46, I-23807 Merate (LC), Italy\\
$^{2}$Universit\`a dell'Insubria, Via Valleggio 11, I-22100 Como, Italy}

\begin{document}
\maketitle
\begin{abstract}
We present a comprehensive spectral-timing study of the  black hole candidate MAXI J1659-152 during its 2010 outburst. We analysed 65 RXTE observations taken along this period and computed the fundamental diagrams commonly used to study black hole transients. We fitted power density and energy spectra and studied the evolution of the spectral and timing parameters along the outburst. We discuss the evolution of the variability observed at different energy bands on the basis of the relative contribution of the disc and hard components to the energy spectrum of the source. We conclude that hard emission accounts for the observed fast variability, it being strongly quenched when type-B oscillations are observed. We find that both disc and hard emission are responsible for local count-rate peaks until the system reaches the soft state. From that point, the peaks are only observed in the hard component, whereas the thermal component drops monotonically probably following the accretion rate decrease.  We have also computed time-lags between soft and hard X-ray variability confirming that lags are larger during the hard-to-soft transition than during the hard state.  \\      
\end{abstract}
\section{Introduction}
Black hole X-ray transients (BHT) spend most of their lives in quiescence, displaying luminosities too low to be detected by X-ray all-sky monitors (see e.g., \citealt{Garcia1998}). They are discovered during outburst events in which their X-ray luminosity increases by several orders of magnitude and their spectral and time variability properties change with time. This leads to the definition of the so-called \textquoteleft states\textquoteright. There is still much discussion about how many different states there are (\citealt{vanderklis2006}; \citealt{Belloni2010} for recent reviews), but X-ray observations have made clear the presence of a $hard$ state (historically known as \textit{low/hard}; LHS) at the beginning of the outburst, which evolves towards a $soft$ state (\textit{high/soft}; HSS). The LHS is also observed at the end of the outburst and it is characterized by a power-law dominated energy spectrum with a power-law index of $\sim 1.6$ (2--20 keV band). This power-law component is though to arise from a \textquoteleft corona\textquoteright of hot electrons, where softer seed photons coming from an accretion disc are up-Comptonized (e.g., \citealt{Gilfanov2010} for a review). Compact radio jets are observed during the LHS (see e.g., \citealt{Fender2006}) and synchrotron emission could also account for the high energy emission during this stage of the outburst (\citealt{Markoff2001}). Aperiodic variability with a fractional root mean square amplitude (rms) above 30\% is also seen. It is almost energy independent (\citealt{Gierli'nski2005}) and sharply correlated with flux (\citealt{Gleissner2004}; \citealt{tmd2011}).\\  
The high energy spectrum softens during HSS since a thermal disc black-body component becomes dominant. The rms drops below 5 per cent and a much more scattered rms-flux correlation is observed (see \citealt{tmd2011} for the evolution of the long term rms-flux relation along the outburst).
The situation is more complex in between these two \textquoteleft canonical\textquoteright ~states. A hard-to-soft transition at high flux is generally observed on relatively short time scales (hours/days) as compared to those seen for the canonical states (weeks/months). During this transition, both timing and spectral properties change dramatically, leading to \textquoteleft intermediate\textquoteright ~states. \cite{Homan2005} and \cite{Belloni2005} identify two additional states, the hard-intermediate state (HIMS) and the soft-intermediate state (SIMS) based on spectral and timing properties (see \citealt{Wijnands1999}, \citealt{Casella2004} and \citealt{Casella2005} for different types of quasi periodic oscillations (QPOs)). In this paper we would follow this classification (see \citealt{McClintock2006} for an alternative classification and \citealt{Motta2009} for a comparison). The count-rate drops considerably during the HSS and a final soft-to-hard transition towards quiescence is usually observed.
Whereas the main properties of the LHS and HSS are known and have been studied in many sources finding a relatively homogeneous behaviour, the study of the whole outburst evolution and state-transitions has proven more elusive and very different behaviours have been reported depending on the system.\\
  
MAXI J1659-152 was discovered  independently by Swift/BAT (GRB 100925A; \citealt{Mangano2010}) 
and MAXI/GSC (\citealt{Negoro2010}) on September 25, 2010. A variable optical counterpart was soon detected (\citealt{Marshall2010}; \citealt{Jelinek2010}; \citealt{Russell2010b}), showing broad, double-peak emission lines ($FWHM\sim2000$ \kms) typical of accreting binaries (\citealt{deUgartePostigo2010b}). The source was detected in radio with a linear polarization level of $\sim 23\%$ (\citealt{vanderHorst2010}), submillimetres (\citealt{deUgartePostigo2010a}) and near infrared (\citealt{D'Avanzo2010}) wavelengths.
At high energies MAXI J1659-152 was also observed by the \textit{Rossi X-ray timing explorer} (RXTE)  and the XMM and INTEGRAL observatories (\citealt{Kuulkers2010a}; \citealt{Vovk2010}). RXTE observations performed 3 days after the discovery revealed strong similarities with the typical timing properties of BHT during the HIMS (\citealt{Kalamkar2010}) indicating that MAXI J1659-152 is a black hole candidate. This was confirmed by the subsequent transitions to the SIMS and HSS observed on October 12 (\citealt{Belloni2010b}) and October 17 (\citealt{Shaposhnikov2010}), respectively. After a short (15 days) stay in soft states, a new transition to the HIMS was observed (\citealt{tmd2010b}).
X-ray dips with a recurrent period of 2.41 hours have been detected in MAXI J1659-152, pointing to a high orbital inclination and suggesting that MAXI J1659-152 is the black hole binary with the shortest orbital period (\citealt{Kuulkers2010b}; \citealt{Belloni2010c}; \citealt{Kuulkers2011}).   
Here, we study in detail the evolution of the spectral and timing properties of the source along the 2010 outburst until observations were interrupted due to Sun constraints. We focus on the evolution of the variability during the  hard-to-soft and soft-to-hard transitions and how it is related to the relative contribution of the various components present in the energy spectra.
\begin{figure*}
\centering
 \includegraphics[width= 14cm,height=10cm]{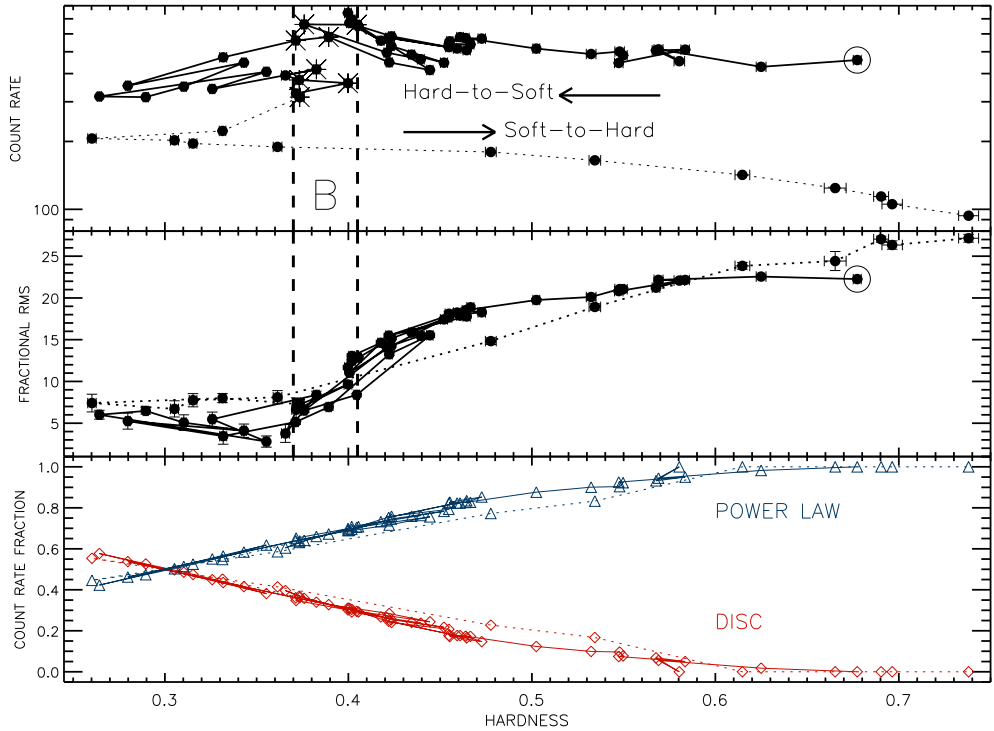}
 \smallskip
\caption{Upper panel: hardness-intensity  diagram obtained using all the RXTE observations available. Intensity correspond to the count rate within the STD2 channels 0--31 (2--15 keV)  and hardness  is defined as the ratio of counts in 11--20 (6.1--10.2 keV) and 4-10 (3.3--6.1 keV) STD2 channels. Each point corresponds to an entire observation. Observations with a star correspond to those with a type-B QPO in the PDS. Solid line joins consecutive observations starting from observation \#1 (big, open circle). Observations taken after the last type-B QPO are joined by a dotted line.  Dashed lines delimit the range in hardness where these oscillation are detected. Middle panel: corresponding hardness-rms diagram within the 0.1--64 Hz frequency band. Lower panel: corresponding power-law (open triangles) and disc (open diamonds) relative contributions to the observed count rate (see Sect. \ref{spectral}).}
\label{hid}
\end{figure*}

\section{Observations}
We analyse 65 RXTE observations of MAXI J1659-152 performed within September 28, 2010 and November 11, 2010.\\ 
The variability study presented in this paper is based on data from the \textit{Proportional Counter Array} (PCA). For some observations the mode GoodXenon1\_2s was used but most of the data are in the mode E\_125us\_64M\_0\_1s, which covers the PCA effective energy range (2-60 keV) with 64 bands. Power density spectra (PDS) for each observation were computed following the procedure outlined in \cite{Belloni2006}. We used stretches 16 s long and PCA channels 0--35 (2--15 keV). 

The PCA Standard 2 mode (STD2) was used for the spectral analysis. It covers the 2--60 keV energy range with 129 channels. From the data, we extracted hardness ($h$), defined as the ratio of counts in STD2 channels 11--20 (6.1--10.2 keV) and 4-10 (3.3--6.1 keV).  
Energy spectra from the PCA 
(background and dead-time corrected) were extracted for each observation using the standard RXTE software within \textsc{heasoft} V. 6.7. For the spectral fitting, Proportional Counter Unit  2 was solely used. In order to account for residual uncertainties in the instrument calibration a systematic error of $0.6\%$ was added to the spectra\footnote{See http://www.universe.nasa.gov/xrays/programs/rxte/pca/doc/rmf/pcarmf-11.7/ for a detailed discussion on the PCA calibration issues.}. %and 

\begin{figure} 
 \includegraphics[width= 8.8cm,height=6.5cm]{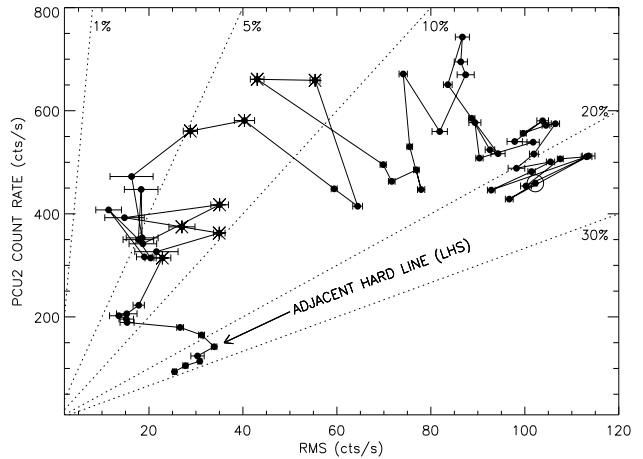}
\caption{Rms-intensity diagram obtained following \citet{tmd2011}. Each point corresponds to an entire observation. A solid line joins consecutive observations starting from observation \#1 (big, open circle). Stars correspond to observations with a type-B QPO in the PDS. Dotted lines represent the 1, 5, 10, 20 and 30 per cent fractional rms levels.}
\label{rid}
\end{figure}

\section{Analysis and results}
We computed the fundamental diagrams commonly used for the study of BHT and performed fits to the energy spectra and PDS. The QPOs present in the PDS have been classified following \cite{Casella2005}. Finally, we have also measured time-lags between soft and hard variability for the only observation long enough to perform this analysis.
\subsection{Fundamental diagrams}
As a first step of the analysis, we computed the hardness-intensity and the hardness-rms diagrams (HID and HRD), which are presented in the upper and middle panels of Fig.\ref{hid}, respectively. The fractional rms was computed within the frequency band 0.1--64 Hz following \cite{Belloni1990}.  We have also computed the rms-intensity diagram (RID) presented in Fig. \ref{rid} following \cite{tmd2011}. Rms values obtained by using a soft (2--6 keV) and a hard (6--15 keV) band are shown in the upper panel of  Fig. \ref{rmsc} as open and filled circles, respectively. The comparison between the rms observed in these two bands is effectively a rms spectrum of two energy bins. This is enough to get a reliable estimation of the energy spectrum of the variability even when the count rate is low. The latter results in large error bars if using narrow energy bands. This method allows us to infer whether the rms spectrum is flat, hard or inverted (i.e. more variability at low energies) for each observation. For a more detailed comparison, we show in Fig. \ref{rmss} three rms spectra corresponding to observations taken along the hard-to-soft transition. They are obtained using six energy bands and give results consistent with those that can be extracted from the upper panel of Fig. \ref{rmsc}.\\ 

The source describes in the HID the standard q-shaped pattern moving from observation \#1 (open, big circle in Fig. \ref{hid}) in the counter clockwise direction. However, the initial flux rise was not observed by RXTE and, as pointed out by \citet{Kalamkar2010}, the first RXTE observation already correspond to the HIMS. This is confirmed by the fact that no hard line (i.e. sharp, linear rms-flux relation; \citealt{tmd2011}) is observed in the RID. After $\sim 16$ days in the HIMS, where the count rate peak is observed, type-B QPOs are seen in the PDS, indicating the system is in the SIMS. Once this state is reached, fast transitions are observed between the SIMS and the HSS. A hard excursion to the HIMS between two soft excursions is observed. After an important decrease in count rate the system reaches the softest (observed) point of the outburst and a soft-to-hard transition is seen. The following can be outlined after a detailed study of the fundamental diagrams:   
\begin{figure*}
\centering
 \includegraphics[width= 14cm,height=10cm]{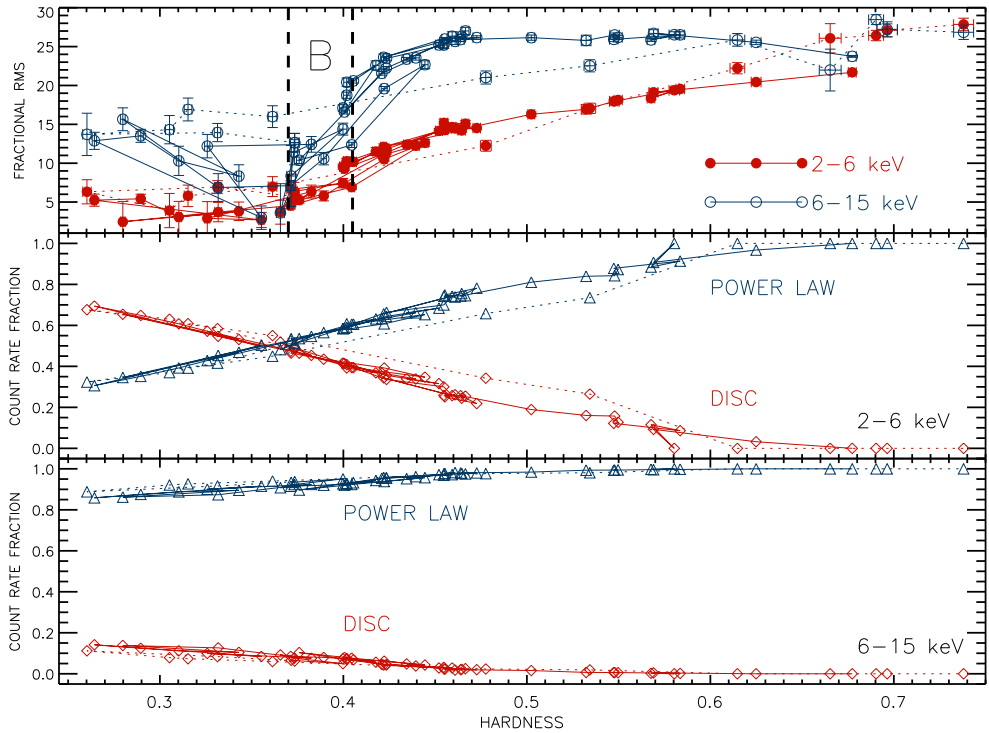}
 \smallskip
\caption{Upper panel: hardness-rms diagrams obtained by using soft (STD2 0-9; 2-6 keV) and hard (STD2 10-31; 6-15 keV) energy bands for computing rms, respectively. Dotted and dashed lines represent the same as in Fig.\ref{hid}. Middle and lower panels: power-law (open triangles) and disc (open diamonds) contribution to the observed count rate within the above bands.}
\label{rmsc}
\end{figure*}
\begin{itemize}
\item The rms decreases monotonically during the first HIMS observations. The corresponding PDS show strong type-C QPOs typical of this state. As usually observed in BHT, a fast decrease in rms is observed during the transition between the HIMS and the SIMS. All the type-B QPOs are observed within $h \sim 0.40$--$0.37$ (see dashed lines in Fig. \ref{hid}) and a minimum in rms is observed at $h \sim 0.35$. From that point, rms increases with softening. This trend is not observed during the second soft excursion and the soft-to-hard transition (dotted line in Fig. \ref{hid}), but a constant rms $\sim 8\%$ value is seen until the system reaches again the HIMS. As far as we know, this clear break in the hardness-rms relation widely observed in BHT has not been reported before (however see discussion in Sect. \ref{discussion}).  The above behaviour is present in both, soft and hard energy bands (Fig. \ref{rmsc}), but it is much more prominent in the hard channels. 

\item During the first observation, the rms shows an almost flat energy spectrum (see upper panel in Fig. \ref{rmss}), consistent with what is observed in early HIMS observations in other systems (see e.g., \citealt{Gierli'nski2005}). Less variability is observed in the soft band as the system gets soft, but in contrast with what is usually observed (see e.g., \citealt{tmd2011} for the case of GX 339-4), the hard rms slightly increases during the HIMS (middle panel in Fig. \ref{rmss}). The rms drops abruptly from the last HIMS observation to the softer SIMS observations. In the hard band, rms fades from $\sim 20\%$ to $4\%$ within the narrow range of hardness where type-B QPOs are observed (dotted lines in Fig. \ref{hid} and Fig. \ref{rmsc}). The rms minimum at  $h \sim 0.35$ is observed in both hard and soft bands, being the rms spectra flat again (see upper panel in Fig. \ref{rmsc} and lower panel in Fig. \ref{rmss}). Between $h \sim 0.25$--$0.35$ we see again much more variability at high energies (upper panel in Fig. \ref{rmsc}). Hard rms spectra are observed during the soft-to-hard transition until they become flat or slightly inverted. This behaviour is typical of the LHS (\citealt{Gierli'nski2005}; \citealt{tmd2010}) and it is observed in the last four RXTE observations.

\item No hard line is observed in the RID (Fig. \ref{rid}). The adjacent hard line seems to be obeyed by the last four RXTE observations, when a flat/inverted rms spectrum are observed, confirming that they correspond to the LHS. We note that this HIMS-LHS transition, which is not obvious when looking at the HID, is sharply marked in the RID. As seen in GX 339-4, type-B QPOs are localized in the $\sim 5$--$10\%$ region of the RID. This $10\%$ line seems to divide precisely HIMS and SIMS observations; as an example, observation  95108-01-21-00 with a very late HIMS PDS (low coherent Type C QPO) has a rms of $11\%$. The $\sim 5\%$ border is not so sharp, especially at count rates $\lesssim 350$ cts s$^{-1}$ where some observations without a type B QPO cross this line. This is due to (i) fast transitions between the HSS and SIMS in the region around $\sim 350$ cts s$^{-1}$, which results in hybrid observations (see also \citealt{tmd2011}), and (ii) the already mentioned break of the usual hardness-rms correlation during the two soft excursions observed. This results in HSS observations with rms $\geqslant 5\%$  and around $\sim 15\%$ in the hard band. Indeed, a clear soft branch (rms $\sim$ 1--5 per cent) is not present in the RID.
\end{itemize}
Finally we show the RXTE light-curve (2--15 keV) during the outburst in Fig. \ref{lc}. SIMS and HSS epochs have been marked with light and dark grey bands, respectively. The thick solid line represents the transition to the LHS.

\subsection{Spectral evolution}
\label{spectral}
We have performed a spectral fitting of the 65 observations analysed in this paper. Energy spectra have been fitted within the band $\sim 4$--22 keV, where RXTE/PCU offers its maximum throughput and spectral calibration is reliable (see e.g., \citealt{Jahoda2006}). We have  used {\sc xspec v 11.3.2}. Given the multitude of spectral models available, we tested several ones in a first approach. This method has been already adopted for other sources (e.g., GX 339-4, \citealt{Nowak2002}; Cyg X-1, \citealt{Wilms2006}).\\
We started with models of one single component, either a cutoff power law or a multicolor disk blackbody. Neither of them could fit the spectra. In order to obtain good fits and acceptable parameters, a model consisting of a simple power-law plus a multi-color disk-blackbody component was used.  
No high energy cut-off associated to the powerlaw component was needed for any of the observations, as expected from the energy range considered ($\leq 22$ keV; e.g., \citealt{Motta2009}, \citealt{Miyakawa2008}). 
A Gaussian emission line with a centroid constrained between 6.4 and 6.8 keV was also needed. A hydrogen column density was used ({\tt wabs} in {\sc xspec}), with N$_{\rm H}$ frozen to $3 \times 10^{21} {\rm cm}^{-2}$, the value derived from {\it Swift}/XRT (\citealt{Kennea2010}). The addition of an iron edge did not significantly improve the fits. 
No evident residuals due to reflection features were evident apart from the iron line, thus no additional reflection component was needed to describe the data. In a second step, we tried to fit the spectra using more sophisticated Comptonization models (\textit{comptt}, \textit{pexrav}) but the result was not statistically better than that obtained by using the model described above. Using \textit{comptt} and \textit{pexrav} we obtain a value of the $\chi_{red}^2$ higher than that obtained with the powerlaw+diskbb model. We conclude that a empirical simple model constituted by a powerlaw+diskbb is sufficient to describe the data.
In Fig. \ref{par} we present the evolution of the main spectral parameters along the outburst evolution. Our results are consistent with a constant inner disc radius\footnote{The normalization for the {\tt diskbb} component is defined as $(\frac{R_{in}/ km}{D/10 kpc})^2 cos \Theta$, where R$_{in}$ is the inner disc radius (km), D is the distance to the source (kpc) and $\Theta$ is the inclination angle of the disk.} around $\sim 40$ km (assuming an orbital inclination of 70 degrees), showing a possible decrease at the end of the outburst. We find an inner disc temperature in the range 0.6-0.9 keV, consistent with the values usually observed in BHT (see e.g., \citealt{Motta2009}). \\ 
The photon index of the power-law component increases from  $\sim 1.9$ during the first HIMS observation to $\sim 2.3$ during the soft states. This value is lower than those usually observed in BHT during soft states. The photon index decreases again during the final soft-to-hard transition where values around  $\sim 1.7$ are reached during the LHS observations. The main spectral parameters obtained for each of the observation are shown in Tab. \ref{tab:spec}.
From this table (see also Fig. \ref{par}) it is clear that our constraints on the disc parameters are sometimes poor. This is expected since by using PCA data we are only able to see the high energy part of the disc black body component above the Wien peak. It is also known that, even if the the {\tt diskbb} model provides a good description of the thermal component, the derived spectral parameters should not be interpreted literally (see e.g. \citealt{Remillard2006b}). However, we note that this thermal component is clearly present in the data and well described by the model we use. Hereafter we focus on the contribution of this thermal component to the total flux rather than in the evolution of single disc parameters to which our study is less sensitive. \\ 
The fractional contribution to the observed count-rate associated with the disc and the power-law component are shown  in the lower panel of Fig. \ref{hid} (2-15 keV), and in the middle and lower panels of Fig. \ref{rmsc} (2-6 keV and 6-15 keV). The following is noted by comparing these results with the hardness and rms evolution:
\begin{itemize}
\item  During the LHS and the HIMS ($h > 0.4$; see Fig. \ref{hid})  the rms is well correlated with the power-law contribution to the total count-rate and anti-correlated with the disc contribution. The fact that within $0.41 \leqslant h \leqslant 0.58$ we see a higher rms during the hard-to-soft transition than during the soft-to-hard (dotted line) can be also explained in terms of power-law contribution to the observed count rate. The same conclusion can be extracted from Fig. \ref{rmsc} when the soft (2-6 keV) and hard bands (6-15 keV) are considered.   
\item During the SIMS ($ 0.37 \leq h \leq 0.40$) the situation is different. The rms decreases drastically, much faster than power-law contribution decreases. This becomes even more evident when we look at the hard band in Fig. \ref{rmsc}, which is clearly power-law dominated. There is no observation within this hardness band during the soft-to-hard transition, but if we consider the closest two observations at both sides of the SIMS it seems probable that during this back transition there is more variability and less power-law contribution than during the hard-to-soft transition. 
\item In the HSS ($h < 0.37$) the behaviour is complex. During the first soft excursion, when count-rate is above $\sim 300$, the rms increases with softening, i.e. with disc contribution. However from Fig. \ref{rmsc} it is clear that in the hard band (power-law dominated) the rms is larger than in the soft band (disc dominated) with the exception of the points very close to the SIMS, where energy spectrum of the rms is flat. During the second soft excursion (count-rate lower than  $\sim 300$) the rms is constant and higher than that observed in the first soft excursion, especially in the points close to the SIMS. The power-law contribution slightly increases as compared to the previous soft excursion and there is much more variability in the hard band than in the soft band. By comparing the two soft excursions it is clear that we see very different variability levels in correspondence with rather similar power-law and disc relative contributions. We note that this second soft excursion at lower count rate and higher rms occurs right after the last type-B QPO is observed.         
\end{itemize} 
\begin{figure}
\includegraphics[width=8.8cm]{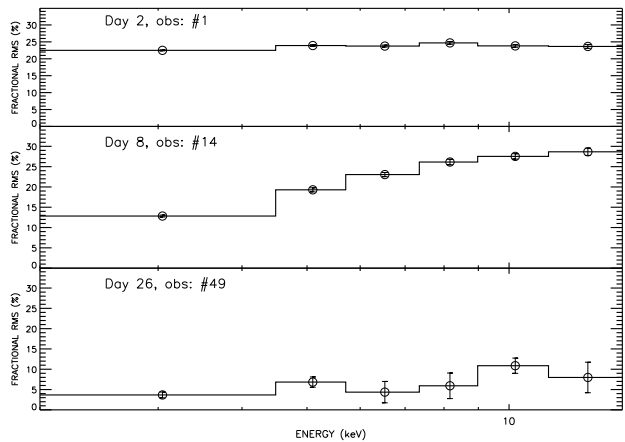}
\caption{Rms spectra calculated for STD2 channels  0--6 (2--4.5 keV), 7--9 (4.5--5.7 keV), 10--13 (5.7--7.3 keV), 14--17 (7.3--9 keV), 18--23 (9--11.4 keV) and 24--31 (11.4--14.8 keV). They cover gradually the behaviour observed during the hard-to-soft transition within the first 49 observations. T$_0$ corresponds to MJD 55465. 
\label{rmss}}
\end{figure}  

\begin{figure}
 \includegraphics[width= 8.8cm,height=6.5cm]{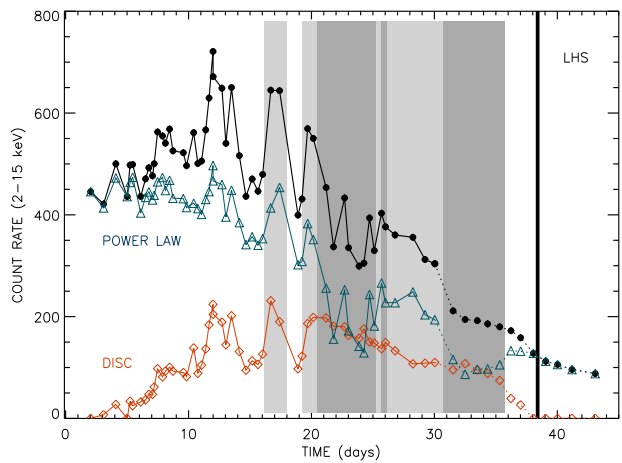}
\caption{Light-curve of the system along the outburst within the band 2-15 keV (STD2 0-31). Count rates associated with the disc and power-law components are shown as open diamonds and open triangles, respectively. SIMS epochs are coloured in light grey, whereas the dark grey regions correspond to the HSS. T$_0$ corresponds to MJD 55465.}
\label{lc}  
\end{figure}
\begin{figure}
 \includegraphics[width= 8.8cm,height=6.5cm]{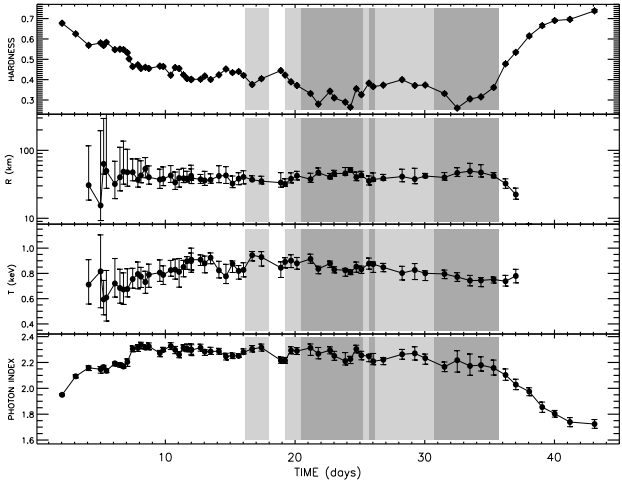}
\caption{Evolution of the hardness and the main spectral parameters during the outburst. $R$ corresponds to the inner disc radius and $T$ to its temperature. SIMS epochs are coloured in light grey, whereas the dark grey regions correspond to the HSS. T$_0$ corresponds to MJD 55465. }
\label{par}
\end{figure}
Using the results from the spectral fits we computed the absolute count-rates associated with the disc and the power-law (2--15 keV). We have over-plotted them in  Fig. \ref{lc}. Only during the HSS epochs (dark grey bands) the disc dominates the observed count-rate. As seen in other systems the disc is not observed within the XTE/PCU band at the beginning and at the end of the outburst (e.g., \citealt{Motta2009}).\\  
We also note that all the wiggles present in the light-curve are observed in both components until the system reaches the HSS for the first time. From that point onwards (day $\sim 21$), the count-rate associated with the disc decreases monotonically, probably following the accretion rate, and the wiggles observed in the light-curve are solely caused by variations in the power-law count rate.   
 
\subsection{Quasi periodic oscillations}
\begin{figure}
 \includegraphics[width= 8.8cm,height=6.5cm]{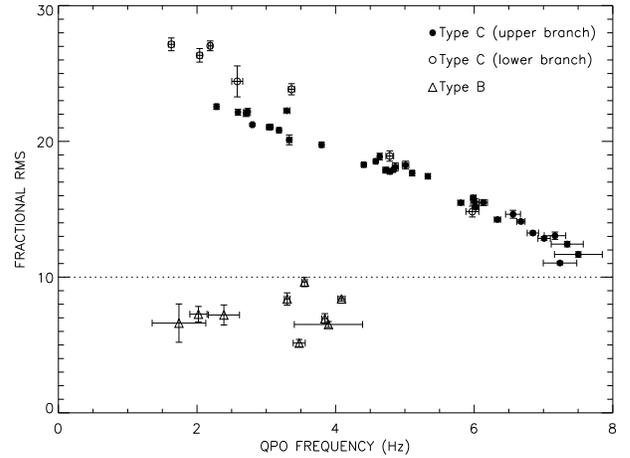}
\caption{Total fractional rms (0.1--64 Hz) as a function of the QPO frequency. Type-C observed during the hard-to-soft and soft-to-hard transitions are marked with a filled  and an open circle, respectively. Open triangles correspond to Type-B QPOs, which do not follow the correlation observed for the type-C and are solely observed when the rms is $\leq 10$ per cent (dotted line). }
\label{qpo}  
\end{figure}

Together with the timing analysis on the evolution of the rms along the outburst and its energy dependence, we have also studied the evolution of the main QPO properties. In table \ref{tab:QPO} we present the fits for the 43 PDS in which Type-C QPOs were observed and the 9 PDS with a type-B QPO. Only in one observation (95118-01-06-00) we see a possible type-A QPO, although its significance is low and we will not consider it in our analysis. PDS fitting was carried out with the standard {\sc xspec} fitting package by using a one-to-one energy-frequency conversion and a unit response. Following \cite{Belloni2002}, we fitted the noise components with three Lorentzians, one zero-centred and other two centred at a few Hz. The QPOs were fitted with one Lorentzian each, only occasionally needing the addition of a Gaussian component to better approximate the shape of the narrow peaks and to reach values of reduced $\chi ^2 $ close to 1. The behaviour of both Type-B and Type-C QPOs is similar to that observed in other BHT (see e.g., \citealt{Belloni2010}). We see the type-C frequency increasing with hardness, whereas type-B are always observed within the frequency range $\sim$ 2--4 Hz.\\ 
Following \cite{Casella2004} and Motta et al. (2011 in prep.) we have plotted total rms as a function of the QPO frequency (Fig. \ref{qpo}). As it was found in those works, Type-C QPOs follow a clear negative correlation. This correlation seems to saturate  around $\sim 7.5$ Hz. Interestingly, if we only consider the Type-C QPOs observed during the soft-to-hard transition (open circles in Fig. \ref{qpo}) a slightly higher slope is observed in the correlation.\\
Type-B QPOs are unequivocally separated from Type-C  using  the rms-frequency representation and they are solely observed when the rms is $\leq 10$ per cent.   
\label{timing}

\subsection{Time-lags}
\begin{figure}
\includegraphics[width=8.5cm]{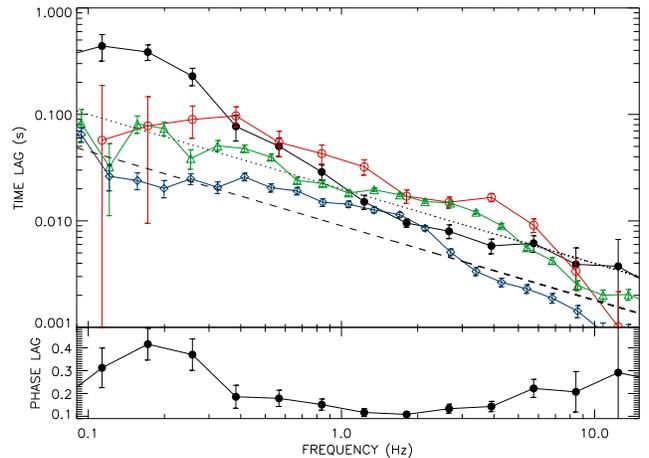}
\caption{Top panel: time-lag vs. frequency for MAXI J1659-152 (black circles). For GX 339-4 we over-plot the time-lags we find during the LHS observation we have analysed (open diamonds) and two HIMS observations at the same hardness (open triangles) and same fracional rms (open circles). The dashed line shows the relation $\Delta t =0.009 \times \nu^{-0.7}$ consistent with what is observed during the LHS. The dotted line shows the relation $\Delta t =0.02 \times \nu^{-0.7}$, which seems more appropriate for HIMS observations.
\label{tlags}}
\end{figure}

Time-lags between soft and hard variability were computed for the first observation (95358-01-02-00), the only one long enough for this purpose. Following \cite{Pottschmidt2000} we used the energy ranges $\sim$ 2--4 keV and  $\sim$ 8--13 keV for the soft and hard bands. They corresponds to the STD2 channels 0--5 and 15--27, respectively. Fast Fourier transforms of each band were computed and cross spectra produced. A positive lag means hard variability lagging soft variability. The obtained time-lags ($\Delta t$) as a function of frequency ($\nu$) are shown in Fig. \ref{tlags} (black circles). As observed in previous works, the time-lag decreases with frequency  within the 0.1-10 Hz band to which we are sensitive. \cite{Pottschmidt2000} noted that in Cyg X-1 the time-lags were larger during the transition between hard and soft states as compared to those observed in the canonical states. The system was in the HIMS during observation \#1 and in agreement with the above work the lags we measure are larger than those typically observed in BHT during LHS. For a direct comparison we have computed time-lags for a LHS observation (diamonds in Fig.\ref{tlags}) and two HIMS observations at the same hardness (triangles in Fig.\ref{tlags}) and same fractional rms (open circles in Fig.\ref{tlags}) corresponding to the 2007 outburst of GX 339-4. Whereas the time-lags corresponding to the LHS observation follows the relation $\Delta t \sim 0.009\times\nu^{-0.7}$ (dashed line in Fig.\ref{tlags}), consistent with the one observed in other systems during that state (see e.g., \citealt{tmd2010} for Cyg X-1 and XTE1752-223) a normalization at least two times bigger ($\sim 0.02$; dashed line in Fig.\ref{tlags}) is needed to account for the time-lags observed in MAXI J1659-152. We note that deviations from the power-law seems to be present at low frequencies.    
 

\section{Discussion} 
\label{discussion}  
The evolution of  MAXI J1659-152 during its 2010 outburst is consistent with that usually observed in black hole transients. The hardness-intensity, rms-hardness and rms-intensity diagrams  are rather typical, with a hard-to-soft transition, a flux decay during a soft (accretion disc dominated) state and a final soft-to-hard transition toward the quiescence. Variability is in general correlated with hardness. Its energy spectrum is hard during the HIMS and the HSS and flat during the LHS.\\ 
If we compare the hardness values with those observed in other systems, we see that MAXI J1659-152 has a rather hard spectrum. This is clear when looking at the spectral parameters that we obtain from the black-body disc + power-law model that fits its energy spectrum. Whereas the disc parameters are within the standard range for a black hole, the power-law index is never higher than $\sim 2.3$~. This can be understood in terms of the high orbital inclination needed to explain the dips observed in the light-curve of the system. As discussed in \cite{Motta2010} for the case of the 2009 ourburst of H 1743-322, assuming that the power-law is arising from a spherical corona and the thermal emission from a thin disc, a hard outburst could be observed in high inclination systems. However, we note that the same system can reach different softening levels (e.g., H1743-322 see also \citealt{McClintock2009})  during different outburst. This shows that other factors, probably related with the available accreting fuel, should play a role. \\       
The rms values observed during some stages of the outburst are similar to those observed in other systems. We see rms close to $\sim 30$ per cent in the LHS,  rms $>$ 10 per cent in the HIMS and $5\% \leq$ rms $\leq 10\%$ in the SIMS. In the HSS we see rms higher than usual, the minimun being $\sim 3 $ per cent. In the softest observation rms is $\sim 8 $ per cent, much higher than usually observed in BHT.  Assuming that much less variability is coming from the disc than from the power-law, a high orbital inclination could in principle explain this behaviour values since we observe less disc photons diluting power-law variability.\\ 
Many of the observed rms values could be explained by solely assuming that the power-law component accounts for the observed variability whereas disc emission varies very little and dilutes power-law variability. Assuming that disc emission does not vary more than the minimum observed rms ($\sim 3 $ per cent) we can roughly recover the rms values we see in the hard states by correcting from the relative contribution of each component along the HIMS.~A similar behaviour was reported by \citet{Shaposhnikov2010a} during the hard-to-soft transition in the BH candidate XTE J1752-223. As discussed by these authors, this can be  understood in terms of variability arising from the Comptonization corona which dominates the LHS, whereas a more stable disc emission starts to contribute significantly to the soft emission during the transition. We note that both a recessing corona resulting in a progressive exposing of the innermost region of the accretion flow, and a truncated disk with an inner radius moving inwards during the transition are able to explain the observations. \\
Variability fades dramatically during the SIMS. Fig.  \ref{rmsc} shows that this is the result of a much less variable power-law emission. We see a jump in the hard  rms (6-15 keV), which drops from $\sim$ 20 to $\sim$ 7 per cent whereas the contribution of the power-law to the total count rate decreases just a little in a monotonic way. This is only observed during the hard-to-soft transition. During the back transition to quiescence the variability level seems much higher even if the power-law contribution is lower. We note that when BHTs cross the SIMS during the hard-to-soft transitions at high flux, relativistic jet ejections are observed (see e.g. \citealt{Fender2006} and \citealt{Fender2009}). As far as we know they have not been seen during the soft-to-hard transition. Thus, the fact that the physical mechanism responsible for power-law variability is removed could be related to the jet ejection. This can be explained if the variability is produced on the base of the jet, since compact radio emission is observed in BHT during the LHS and the HIMS. However, jet emission should not be present in the HSS, when we clearly see power-law variability.  More than one variability component is needed for that model to work.\\       
The evolution of the rms during the HSS is at odds with what is usually observed. In the HRD we see fractional rms increasing with softening at high count rate (first soft excursion) and hardness independent rms at lower fluxes (second soft excursion). The minimum in rms is observed close to the SIMS, and power-law and disc variabilities are consistent with $\sim$ 3 per cent fractional rms. Thus, the $\sim 20$--$25$ per cent power-law variability observed at the end of the HIMS almost disappear after crossing the SIMS. This cannot be explained in terms of an increase of the disc contribution. Indeed, the RID shows that absolute (i.e. non fractional) rms fades dramatically from the HIMS to the HSS. Power-law variability is recovered when the system softens and especially when the total count rate drops after the last type-B QPO is observed. If we associate type-B QPOs with jet ejections  (e.g., \citealt{Soleri2008};\citealt{Fender2009}), we see rms increasing when the system abandons the region where they are observed and when the type-B/jet mechanism is suppressed. \cite{Belloni2005} observed a similar behaviour during the hard-to-soft transition in GX 339-4 (see also \citealt{Fender2009} for XTE J1859+226 and XTE J1550-564). They find similar (2--3$\%$) variability levels in observations showing type-A QPOs, which have hardness values softer but close to those observed in type-B observations. The rms rises for a while after the observations with a type-A and fades again during the softest observations. GX 339-4 crosses the same region with higher rms during the soft-hard-transition. In MAXI J1659-152 we see a possible type-A QPO only in one observation, although its significance is low and it is observed in one of the softest observations (i.e. with higher rms). However, statistics are much lower in this case than for GX 339-4 and it could be the case that we are not sensitive enough to detect those type-A or that they are not present for other reasons. Assuming that these low rms observations of MAXI J1659-152 after the type-B region correspond to those with type-A in GX 339-4, the behaviour of the two sources is similar with the exception of the final softening and variability decrease observed in GX 339-4 (see Fig. 10 in \citealt{Belloni2005}). The latter can be explained by the fact that a strong disc dominated soft state is missing in MAXI J1659-152. As discussed above, this can be understood in terms of the high orbital inclination of the system. In addition or alternatively to this, we should take into account that the orbital period proposed for this system (\citealt{Kuulkers2010b}; \citealt{Belloni2010c}) is much sorter than the one of GX 339-4. This should favour a shorter outburst (as observed) and it could have also an effect on the properties of the steady optically thick accretion flow expected to dominate during soft states.\\  
Only type-C and type-B QPOs are observed in MAXI J1659-152. Apart from their intrinsic differences noted in \cite{Casella2005}, they are clearly separated using a frequency-fractional rms representation. Whereas type-C follow a clear correlation, type-B are clearly outside this correlation, showing a roughly constant rms as function of the frequency. This has been observed by \cite{Casella2004} and  Motta et al. (2011 in prep.) for the cases of XTE J1859+226 and GX 339-4, respectively.\\  
Jet ejections are known to be connected to flux peaks (\citealt{Fender2004}) and to HIMS/SIMS transitions (i.e. type-B QPOs \citealt{Soleri2008}; \citealt{Fender2009}).  There are known exceptions, as the case of the strong X-ray flux peak of the 1998/1999 outburst of XTE J1550-564, where a jet ejection $\sim $4 days after the peak was observed before the transition to the soft states (\citealt{Hannikainen2001}; \citealt{Corbel2002}). In MAXI J1659-152,  type-B QPOs are found in correspondence with local count rate peaks (see light grey bands in Fig. \ref{lc}), especially at high fluxes.  However, the absolute count rate peak is observed before the transition to the SIMS (Fig. \ref{lc}) and from that point radio emission is observed to quench (\citealt{vanderHorst2010b}). No QPO is observed in that observation (either type-B or C) and the rms is that expected for the HIMS. Similar behaviour is seen in the already mentioned very bright observation of XTE J1550-564, although weak QPOs are present in this case. We expect that multi-wavelength studies performed during this outburst of MAXI J1659-152  will be able to discuss our results in light of the detection or non-detection of relativistic jet emission during the observed X-ray count rate peaks. A possible scenario in which both, jet ejections and HIMS/SIMS transitions are related to flux peaks but not between each others cannot be ruled out.\\
It is also remarkable that once the system reaches the HSS all the wiggles and flux peaks present in the light-curve seem related to variations in the power-law emission whereas the disc emission decreases monotonically. This is not observed before the first HSS observation and it is probably connected with the formation of a steady accretion disc during the HSS. Our study suggests that all the scatter usually observed during HSS in the hard-intensity diagrams  and rms-intensity diagrams of BHT (see e.g., \citealt{Dunn2010}) is due to a change in the relative contribution of the power-law, whereas disc emission drops monotonically probably following the accretion rate decrease.\\

We have measured time-lags between soft and hard emission finding a similar frequency dependence time-lag distribution than that already observed in other BHT during hard states (\citealt{Nowak1999}; \citealt{Pottschmidt2000}; \citealt{tmd2010}). The observation in which it was possible to perform this study belongs to the HIMS, and by comparing with the case of GX 339-4 we obtain results in agreement with those reported by \cite{Pottschmidt2000} in Cyg X-1. In the hard-to-soft transition lags are higher than in the LHS.  As discussed in the above works and extensively in \citet{Nowak1999a}, both the time-lags measured in several systems and the observed time-lag evolution cannot be explained by purely Comptonization models, and alternative scenarios (see e.g., \citealt{Kazanas1997}; \citealt{Kording2004}) should be further explored.\\
Finally, we note that by the time we submitted the last revised version of this manuscript another paper based on RXTE analysis on the same source has been accepted for publication (\citealt{kalamkar2011}). It is focussed on identifying the black hole nature of the source and according to arXiv was submitted about the same date than this work. 
      
\section{Conclusions}
We have performed an X-ray spectral and timing analysis of the black hole candidate MAXI J1659-152 during its first observed outburst. The outburst evolution of the system is similar to that previously observed in other black hole candidates, although it presents clear peculiarities especially in what regards to the evolution of the fast variability and its energy dependence. We have discussed this behaviour on the basis of the spectral decomposition we have performed. Complementary results obtained through multi-wavelengths campaigns of the present and forthcoming outbursts of this source will result in a deeper understanding of the behaviour observed in this source and of the accretion process taken place in black hole binaries.   
\vspace{1cm}

\noindent The research leading to these results has received funding from the European Community's Seventh Framework Programme (FP7/2007-2013) under grant agreement number ITN 215212 \textquotedblleft Black Hole Universe\textquotedblright. SM and TB acknowledge support to the ASI grant I/088/06/0.

\bibliographystyle{mn2e}
\bibliography{/Users/tmd/Documents/Liberia_RX.bib}

%%_____________________BEGIN__________TABLE_2____________________________%%
%\begin{table*}																																
%\renewcommand{\arraystretch}{1.3}																																
\onecolumn																																
\begin{center}																																
\begin{longtable}{|c|c|c|c|c|c|c|c|c|}																																
\caption{Spectral parameters derived from the best fit for each observation. A model consisting of a power-law, a multi-color disk-blackbody and a Gaussian emission line with a centroid constrained between 6.4 and 6.8 keV was used. Fluxes (erg s$^{-1}$ cm$^{-2}$) are computed within the band 4-22 keV. }\label{tab:spec}\\																																
\endfirsthead																																
\multicolumn{9}{c}%																																						
{{\tablename\ \thetable{} -- continued from previous page}} \\ \hline																																						
Obs. Num.	&	MJD	&	Obs. ID	&	$\chi ^2_{red} $	&		kT	(keV)						&	R	(km)							&	$\Gamma$								&	Disc flux  &	 Powerlaw flux\\
\hline																																						
\hline																																						
\endhead																																						
\hline \multicolumn{9}{c}{{Continued on next page}} \\																																						
\endfoot																																						
\hline																																						
\endlastfoot																																						
\hline																																						
Obs. Num.	&	MJD	&	Obs. ID	&	$\chi ^2_{red} $	&		kT	(keV)						&	R	(km)							&	$\Gamma$								&	Disc flux &	Powerlaw flux\\
				\hline																																		
				\hline

1	&	55467.0	&	95358-01-02-00	&	0.66	&	$	 -						$	&	$	31	_{-	12	}^{+	86	}	$	&	$	1.95	_{-	0.01	}^{+	0.01	}	$	&	0.00E+00	&	4.87E-09	\\
2	&	55468.1	&	95358-01-02-01	&	1.57	&	$	0.40	_{-	0.40	}^{+	2.55	}	$	&	$	15	_{-	6	}^{+	179	}	$	&	$	2.09	_{-	0.01	}^{+	0.01	}	$	&	5.28E-12	&	4.26E-09	\\
3	&	55469.1	&	95358-01-02-02	&	0.61	&	$	0.71	_{-	0.15	}^{+	0.20	}	$	&	$	63	_{-	19	}^{+	232	}	$	&	$	2.16	_{-	0.02	}^{+	0.01	}	$	&	7.50E-11	&	4.72E-09	\\
4	&	55470.0	&	95108-01-01-00	&	1.01	&	$	0.82	_{-	0.29	}^{+	0.29	}	$	&	$	50	_{-	22	}^{+	303	}	$	&	$	2.15	_{-	0.03	}^{+	0.02	}	$	&	7.12E-11	&	4.29E-09	\\
5	&	55470.2	&	95358-01-03-00	&	0.82	&	$	0.59	_{-	0.12	}^{+	0.15	}	$	&	$	32	_{-	12	}^{+	37	}	$	&	$	2.16	_{-	0.01	}^{+	0.01	}	$	&	6.73E-11	&	4.62E-09	\\
6	&	55470.5	&	95108-01-02-00	&	0.84	&	$	0.61	_{-	0.19	}^{+	0.21	}	$	&	$	40	_{-	15	}^{+	71	}	$	&	$	2.14	_{-	0.02	}^{+	0.01	}	$	&	5.18E-11	&	4.78E-09	\\
7	&	55471.1	&	95358-01-03-01	&	0.60	&	$	0.72	_{-	0.11	}^{+	0.20	}	$	&	$	49	_{-	17	}^{+	88	}	$	&	$	2.19	_{-	0.02	}^{+	0.01	}	$	&	8.99E-11	&	3.97E-09	\\
8	&	55471.5	&	95108-01-03-00	&	1.08	&	$	0.68	_{-	0.12	}^{+	0.15	}	$	&	$	48	_{-	18	}^{+	56	}	$	&	$	2.18	_{-	0.02	}^{+	0.02	}	$	&	9.28E-11	&	4.29E-09	\\
9	&	55471.8	&	95108-01-04-00	&	1.03	&	$	0.67	_{-	0.12	}^{+	0.13	}	$	&	$	48	_{-	12	}^{+	27	}	$	&	$	2.17	_{-	0.01	}^{+	0.02	}	$	&	1.18E-10	&	4.42E-09	\\
10	&	55472.1	&	95108-01-05-00	&	0.84	&	$	0.68	_{-	0.06	}^{+	0.16	}	$	&	$	37	_{-	8	}^{+	38	}	$	&	$	2.20	_{-	0.03	}^{+	0.02	}	$	&	1.18E-10	&	4.20E-09	\\
11	&	55472.2	&	95358-01-03-02	&	1.05	&	$	0.73	_{-	0.04	}^{+	0.04	}	$	&	$	40	_{-	6	}^{+	9	}	$	&	$	2.26	_{-	0.01	}^{+	0.01	}	$	&	1.96E-10	&	4.20E-09	\\
12	&	55472.5	&	95108-01-06-00	&	0.49	&	$	0.76	_{-	0.07	}^{+	0.09	}	$	&	$	43	_{-	9	}^{+	28	}	$	&	$	2.31	_{-	0.02	}^{+	0.02	}	$	&	2.90E-10	&	4.34E-09	\\
13	&	55472.9	&	95108-01-07-00	&	0.81	&	$	0.80	_{-	0.12	}^{+	0.10	}	$	&	$	53	_{-	14	}^{+	25	}	$	&	$	2.31	_{-	0.03	}^{+	0.02	}	$	&	2.60E-10	&	4.40E-09	\\
14	&	55473.1	&	95108-01-08-00	&	0.92	&	$	0.77	_{-	0.09	}^{+	0.07	}	$	&	$	40	_{-	8	}^{+	19	}	$	&	$	2.34	_{-	0.03	}^{+	0.02	}	$	&	2.86E-10	&	4.12E-09	\\
15	&	55473.5	&	95108-01-09-00	&	0.82	&	$	0.73	_{-	0.09	}^{+	0.09	}	$	&	$	37	_{-	10	}^{+	16	}	$	&	$	2.31	_{-	0.01	}^{+	0.02	}	$	&	2.84E-10	&	4.35E-09	\\
16	&	55473.7	&	95108-01-10-00	&	0.90	&	$	0.79	_{-	0.10	}^{+	0.08	}	$	&	$	38	_{-	8	}^{+	19	}	$	&	$	2.33	_{-	0.03	}^{+	0.03	}	$	&	2.94E-10	&	4.00E-09	\\
17	&	55474.6	&	95108-01-11-00	&	0.63	&	$	0.81	_{-	0.07	}^{+	0.10	}	$	&	$	43	_{-	9	}^{+	17	}	$	&	$	2.27	_{-	0.03	}^{+	0.02	}	$	&	2.93E-10	&	4.09E-09	\\
18	&	55474.8	&	95108-01-12-00	&	1.00	&	$	0.79	_{-	0.08	}^{+	0.07	}	$	&	$	34	_{-	7	}^{+	14	}	$	&	$	2.30	_{-	0.01	}^{+	0.02	}	$	&	2.59E-10	&	3.88E-09	\\
19	&	55475.4	&	95108-01-13-00	&	0.59	&	$	0.83	_{-	0.05	}^{+	0.08	}	$	&	$	39	_{-	10	}^{+	13	}	$	&	$	2.33	_{-	0.03	}^{+	0.03	}	$	&	4.65E-10	&	3.91E-09	\\
20	&	55475.8	&	95108-01-14-00	&	0.58	&	$	0.83	_{-	0.07	}^{+	0.08	}	$	&	$	39	_{-	5	}^{+	16	}	$	&	$	2.30	_{-	0.03	}^{+	0.03	}	$	&	2.99E-10	&	3.86E-09	\\
21	&	55476.1	&	95108-01-15-00	&	0.91	&	$	0.81	_{-	0.12	}^{+	0.11	}	$	&	$	38	_{-	4	}^{+	11	}	$	&	$	2.26	_{-	0.02	}^{+	0.02	}	$	&	3.44E-10	&	3.82E-09	\\
22	&	55476.4	&	95108-01-16-00	&	0.60	&	$	0.85	_{-	0.08	}^{+	0.05	}	$	&	$	42	_{-	6	}^{+	18	}	$	&	$	2.31	_{-	0.02	}^{+	0.02	}	$	&	4.75E-10	&	4.00E-09	\\
23	&	55476.7	&	95108-01-17-00	&	0.85	&	$	0.90	_{-	0.06	}^{+	0.04	}	$	&	$	39	_{-	7	}^{+	14	}	$	&	$	2.31	_{-	0.02	}^{+	0.03	}	$	&	6.82E-10	&	4.15E-09	\\
24	&	55477.0	&	95108-01-18-00	&	1.31	&	$	0.90	_{-	0.09	}^{+	0.06	}	$	&	$	37	_{-	5	}^{+	7	}	$	&	$	2.30	_{-	0.04	}^{+	0.04	}	$	&	8.33E-10	&	4.64E-09	\\
25	&	55477.0	&	95108-01-18-01	&	0.91	&	$	0.91	_{-	0.08	}^{+	0.09	}	$	&	$	36	_{-	6	}^{+	13	}	$	&	$	2.30	_{-	0.04	}^{+	0.04	}	$	&	7.71E-10	&	4.37E-09	\\
26	&	55477.7	&	95108-01-19-00	&	0.97	&	$	0.91	_{-	0.05	}^{+	0.04	}	$	&	$	37	_{-	4	}^{+	8	}	$	&	$	2.31	_{-	0.03	}^{+	0.03	}	$	&	7.14E-10	&	4.27E-09	\\
27	&	55478.0	&	95108-01-20-00	&	1.05	&	$	0.88	_{-	0.08	}^{+	0.06	}	$	&	$	42	_{-	9	}^{+	18	}	$	&	$	2.28	_{-	0.03	}^{+	0.02	}	$	&	5.25E-10	&	3.73E-09	\\
28	&	55478.5	&	95108-01-21-00	&	0.92	&	$	0.92	_{-	0.05	}^{+	0.04	}	$	&	$	43	_{-	10	}^{+	15	}	$	&	$	2.29	_{-	0.02	}^{+	0.03	}	$	&	7.76E-10	&	4.22E-09	\\
29	&	55479.1	&	95108-01-22-00	&	0.66	&	$	0.82	_{-	0.06	}^{+	0.07	}	$	&	$	32	_{-	4	}^{+	10	}	$	&	$	2.29	_{-	0.02	}^{+	0.02	}	$	&	4.39E-10	&	3.62E-09	\\
30	&	55479.7	&	95108-01-23-00	&	0.62	&	$	0.78	_{-	0.06	}^{+	0.09	}	$	&	$	38	_{-	7	}^{+	9	}	$	&	$	2.25	_{-	0.03	}^{+	0.02	}	$	&	2.93E-10	&	3.27E-09	\\
31	&	55480.2	&	95108-01-24-00	&	0.72	&	$	0.88	_{-	0.07	}^{+	0.02	}	$	&	$	41	_{-	8	}^{+	10	}	$	&	$	2.25	_{-	0.02	}^{+	0.02	}	$	&	4.09E-10	&	3.41E-09	\\
32	&	55480.7	&	95108-01-25-00	&	1.41	&	$	0.82	_{-	0.06	}^{+	0.04	}	$	&	$	37	_{-	1	}^{+	5	}	$	&	$	2.25	_{-	0.01	}^{+	0.02	}	$	&	3.52E-10	&	3.26E-09	\\
33	&	55481.0	&	95108-01-26-00	&	1.30	&	$	0.83	_{-	0.05	}^{+	0.06	}	$	&	$	35	_{-	3	}^{+	6	}	$	&	$	2.28	_{-	0.01	}^{+	0.03	}	$	&	4.24E-10	&	3.33E-09	\\
34	&	55481.7	&	95108-01-27-00	&	0.77	&	$	0.94	_{-	0.05	}^{+	0.03	}	$	&	$	33	_{-	5	}^{+	14	}	$	&	$	2.30	_{-	0.02	}^{+	0.03	}	$	&	9.11E-10	&	3.86E-09	\\
35	&	55482.4	&	95108-01-28-00	&	0.83	&	$	0.93	_{-	0.07	}^{+	0.04	}	$	&	$	32	_{-	2	}^{+	6	}	$	&	$	2.32	_{-	0.03	}^{+	0.03	}	$	&	7.35E-10	&	4.21E-09	\\
36	&	55483.9	&	95108-01-30-00	&	0.91	&	$	0.84	_{-	0.08	}^{+	0.05	}	$	&	$	38	_{-	5	}^{+	9	}	$	&	$	2.22	_{-	0.02	}^{+	0.03	}	$	&	3.36E-10	&	2.94E-09	\\
37	&	55484.2	&	95118-01-01-00	&	0.45	&	$	0.89	_{-	0.05	}^{+	0.03	}	$	&	$	42	_{-	6	}^{+	7	}	$	&	$	2.21	_{-	0.02	}^{+	0.02	}	$	&	4.52E-10	&	3.00E-09	\\
38	&	55484.7	&	95118-01-01-01	&	0.81	&	$	0.90	_{-	0.06	}^{+	0.05	}	$	&	$	38	_{-	4	}^{+	8	}	$	&	$	2.29	_{-	0.03	}^{+	0.03	}	$	&	6.97E-10	&	3.59E-09	\\
39	&	55485.1	&	95118-01-02-00	&	0.67	&	$	0.88	_{-	0.05	}^{+	0.05	}	$	&	$	47	_{-	4	}^{+	8	}	$	&	$	2.29	_{-	0.03	}^{+	0.03	}	$	&	7.21E-10	&	3.31E-09	\\
40	&	55486.2	&	95118-01-03-01	&	0.59	&	$	0.91	_{-	0.03	}^{+	0.04	}	$	&	$	41	_{-	2	}^{+	5	}	$	&	$	2.31	_{-	0.06	}^{+	0.03	}	$	&	7.50E-10	&	2.38E-09	\\
41	&	55486.8	&	95118-01-03-00	&	0.93	&	$	0.84	_{-	0.04	}^{+	0.02	}	$	&	$	46	_{-	4	}^{+	5	}	$	&	$	2.27	_{-	0.04	}^{+	0.05	}	$	&	6.18E-10	&	1.48E-09	\\
42	&	55487.7	&	95118-01-04-00	&	0.88	&	$	0.88	_{-	0.03	}^{+	0.02	}	$	&	$	46	_{-	3	}^{+	10	}	$	&	$	2.29	_{-	0.03	}^{+	0.03	}	$	&	6.52E-10	&	2.37E-09	\\
43	&	55488.0	&	95118-01-05-00	&	1.08	&	$	0.83	_{-	0.03	}^{+	0.02	}	$	&	$	51	_{-	4	}^{+	4	}	$	&	$	2.25	_{-	0.03	}^{+	0.04	}	$	&	5.51E-10	&	1.65E-09	\\
44	&	55488.9	&	95118-01-05-01	&	0.72	&	$	0.82	_{-	0.05	}^{+	0.01	}	$	&	$	40	_{-	5	}^{+	8	}	$	&	$	2.21	_{-	0.03	}^{+	0.07	}	$	&	5.27E-10	&	1.38E-09	\\
45	&	55489.3	&	95118-01-06-00	&	1.20	&	$	0.81	_{-	0.02	}^{+	0.02	}	$	&	$	43	_{-	3	}^{+	6	}	$	&	$	2.23	_{-	0.03	}^{+	0.02	}	$	&	5.75E-10	&	1.25E-09	\\
46	&	55489.7	&	95118-01-06-01	&	1.26	&	$	0.85	_{-	0.04	}^{+	0.04	}	$	&	$	35	_{-	4	}^{+	7	}	$	&	$	2.30	_{-	0.03	}^{+	0.03	}	$	&	5.25E-10	&	2.27E-09	\\
47	&	55490.1	&	95118-01-07-01	&	1.12	&	$	0.83	_{-	0.03	}^{+	0.03	}	$	&	$	37	_{-	2	}^{+	9	}	$	&	$	2.26	_{-	0.04	}^{+	0.03	}	$	&	5.02E-10	&	1.73E-09	\\
48	&	55490.7	&	95118-01-07-00	&	1.33	&	$	0.88	_{-	0.04	}^{+	0.04	}	$	&	$	39	_{-	2	}^{+	6	}	$	&	$	2.25	_{-	0.04	}^{+	0.01	}	$	&	4.95E-10	&	2.55E-09	\\
49	&	55491.0	&	95118-01-08-00	&	0.50	&	$	0.87	_{-	0.07	}^{+	0.02	}	$	&	$	41	_{-	7	}^{+	12	}	$	&	$	2.21	_{-	0.03	}^{+	0.06	}	$	&	5.36E-10	&	2.22E-09	\\
50	&	55491.8	&	95118-01-09-00	&	0.98	&	$	0.85	_{-	0.03	}^{+	0.03	}	$	&	$	38	_{-	5	}^{+	17	}	$	&	$	2.22	_{-	0.04	}^{+	0.02	}	$	&	4.62E-10	&	2.21E-09	\\
51	&	55493.3	&	95118-01-10-00	&	1.78	&	$	0.80	_{-	0.05	}^{+	0.05	}	$	&	$	42	_{-	2	}^{+	4	}	$	&	$	2.26	_{-	0.04	}^{+	0.04	}	$	&	3.47E-10	&	2.37E-09	\\
52	&	55494.2	&	95118-01-11-00	&	0.85	&	$	0.83	_{-	0.08	}^{+	0.05	}	$	&	$	40	_{-	4	}^{+	7	}	$	&	$	2.27	_{-	0.05	}^{+	0.05	}	$	&	3.65E-10	&	1.93E-09	\\
53	&	55495.0	&	95118-01-12-00	&	0.62	&	$	0.80	_{-	0.06	}^{+	0.02	}	$	&	$	47	_{-	5	}^{+	9	}	$	&	$	2.23	_{-	0.05	}^{+	0.03	}	$	&	3.54E-10	&	1.87E-09	\\
54	&	55496.5	&	95118-01-13-00	&	0.86	&	$	0.80	_{-	0.04	}^{+	0.03	}	$	&	$	49	_{-	8	}^{+	15	}	$	&	$	2.17	_{-	0.04	}^{+	0.04	}	$	&	3.05E-10	&	1.15E-09	\\
55	&	55497.5	&	95118-01-14-00	&	0.76	&	$	0.77	_{-	0.04	}^{+	0.03	}	$	&	$	47	_{-	7	}^{+	14	}	$	&	$	2.22	_{-	0.09	}^{+	0.07	}	$	&	3.30E-10	&	8.43E-10	\\
56	&	55498.5	&	95118-01-15-00	&	0.58	&	$	0.74	_{-	0.04	}^{+	0.04	}	$	&	$	42	_{-	3	}^{+	5	}	$	&	$	2.17	_{-	0.07	}^{+	0.09	}	$	&	2.77E-10	&	9.58E-10	\\
57	&	55499.3	&	95118-01-15-01	&	0.57	&	$	0.74	_{-	0.05	}^{+	0.02	}	$	&	$	32	_{-	5	}^{+	6	}	$	&	$	2.18	_{-	0.05	}^{+	0.06	}	$	&	2.57E-10	&	9.65E-10	\\
58	&	55500.3	&	95118-01-16-00	&	0.77	&	$	0.75	_{-	0.03	}^{+	0.02	}	$	&	$	22	_{-	3	}^{+	5	}	$	&	$	2.16	_{-	0.05	}^{+	0.06	}	$	&	2.20E-10	&	1.05E-09	\\
59	&	55501.2	&	95118-01-16-01	&	0.87	&	$	0.74	_{-	0.04	}^{+	0.05	}	$	&	$	37	_{-	14	}^{+	138	}	$	&	$	2.10	_{-	0.04	}^{+	0.05	}	$	&	1.13E-10	&	1.36E-09	\\
60	&	55502.0	&	95118-01-17-00	&	0.90	&	$	0.78	_{-	0.05	}^{+	0.05	}	$	&	$	37	_{-	14	}^{+	138	}	$	&	$	2.03	_{-	0.04	}^{+	0.04	}	$	&	8.23E-11	&	1.40E-09	\\
61	&	55503.1	&	95118-01-17-01	&	0.99	&	$	 -						$	&	$	37	_{-	14	}^{+	138	}	$	&	$	1.97	_{-	0.03	}^{+	0.03	}	$	&	0.00E+00	&	1.39E-09	\\
62	&	55504.1	&	95118-01-18-00	&	0.80	&	$	 -						$	&	$	37	_{-	14	}^{+	138	}	$	&	$	1.85	_{-	0.04	}^{+	0.04	}	$	&	0.00E+00	&	1.29E-09	\\
63	&	55505.0	&	95118-01-19-00	&	1.48	&	$	 -						$	&	$	37	_{-	14	}^{+	138	}	$	&	$	1.80	_{-	0.03	}^{+	0.02	}	$	&	0.00E+00	&	1.25E-09	\\
64	&	55506.2	&	95118-01-20-00	&	1.22	&	$	 -						$	&	$	37	_{-	14	}^{+	138	}	$	&	$	1.74	_{-	0.04	}^{+	0.03	}	$	&	0.00E+00	&	1.17E-09	\\
65	&	55508.1	&	95118-01-21-00	&	0.85	&	$	 -						$	&	$	37	_{-	14	}^{+	138	}	$	&	$	1.72	_{-	0.03	}^{+	0.03	}	$	&	0.00E+00	&	1.08E-09	\\

\end{longtable}
\end{center}
%\end{table*}
\twocolumn
%%_____________________PDS_TABLE____________________________%%																														
%\begin{table*}																														
%\renewcommand{\arraystretch}{1.3}																														
\onecolumn																														
\begin{center}																														
\begin{longtable}{|c|c|c|c|c|c|c|}																														
\caption{Best fit for the central peak of the QPO detected during the whole outburst.  The noise components were fitted with three Lorentzian shapes, one zero-centred and other two centred at a few Hz. The QPOs were fitted with one Lorentzian each. Rms is the total, fractional rms (i.e. all the PDS components) within the band 0.1-64 Hz.}\label{tab:QPO}\\																														
\endfirsthead																														
\multicolumn{5}{c}%																														
{{\tablename\ \thetable{} -- continued from previous page}} \\ \hline																														
Observation ID	 &		Frequency	(Hz)			&	Width (Hz)				&		Normalization				&		rms ($\%$)		\\
\hline																														
\hline																														
\endhead																														
\hline \multicolumn{5}{c}{{Continued on next page}} \\																														
\endfoot																														
\hline																															
\endlastfoot																															
\hline																															
Observation ID	&		Frequency	(Hz)			&	Width (Hz)				&		Normalization				&		rms ($\%$)	\\
\hline																															
\hline

95358-01-02-00	&	$	3.30	\pm	0.04	$	&	$	0.53	\pm	0.32	$	&	$	1	\pm	1	$	&	$	22.3	\pm	0.1	$	\\
95358-01-02-01	&	$	2.28	\pm	0.02	$	&	$	0.38	\pm	0.07	$	&	$	14	\pm	1	$	&	$	22.6	\pm	0.2	$	\\
95358-01-02-02	&	$	5.5	\pm	0.2	$	&	$	0.77	\pm	0.61	$	&	$	1.3	\pm	0.7	$	&	$	22.2	\pm	0.3	$	\\
95108-01-01-00	&	$	2.71	\pm	0.03	$	&	$	0.32	\pm	0.08	$	&	$	11	\pm	2	$	&	$	22.1	\pm	0.3	$	\\
95358-01-03-00	&	$	2.80	\pm	0.02	$	&	$	0.34	\pm	0.05	$	&	$	11.6	\pm	1.2	$	&	$	21.2	\pm	0.1	$	\\
95108-01-02-00	&	$	5.3	\pm	0.1	$	&	$	0.61	\pm	0.36	$	&	$	1.0	\pm	0.6	$	&	$	22.2	\pm	0.2	$	\\
95358-01-03-01	&	$	3.18	\pm	0.02	$	&	$	0.41	\pm	0.09	$	&	$	11	\pm	2	$	&	$	20.8	\pm	0.2	$	\\
95108-01-03-00	&	$	3.04	\pm	0.04	$	&	$	0.37	\pm	0.15	$	&	$	11	\pm	4	$	&	$	21.1	\pm	0.2	$	\\
95108-01-04-00	&	$	3.06	\pm	0.03	$	&	$	0.34	\pm	0.09	$	&	$	25	\pm	5	$	&	$	21.1	\pm	0.2	$	\\
95108-01-05-00	&	$	3.33	\pm	0.04	$	&	$	0.32	\pm	0.16	$	&	$	5	\pm	2	$	&	$	20.1	\pm	0.4	$	\\
95358-01-03-02	&	$	3.80	\pm	0.02	$	&	$	0.45	\pm	0.07	$	&	$	5.8	\pm	0.7	$	&	$	19.8	\pm	0.2	$	\\
95108-01-06-00	&	$	4.58	\pm	0.03	$	&	$	0.50	\pm	0.08	$	&	$	11	\pm	1	$	&	$	18.5	\pm	0.2	$	\\
95108-01-07-00	&	$	4.41	\pm	0.03	$	&	$	0.54	\pm	0.10	$	&	$	18	\pm	2	$	&	$	18.3	\pm	0.1	$	\\
95108-01-08-00	&	$	4.84	\pm	0.02	$	&	$	0.54	\pm	0.07	$	&	$	9.0	\pm	0.9	$	&	$	17.9	\pm	0.1	$	\\
95108-01-09-00	&	$	4.72	\pm	0.04	$	&	$	0.53	\pm	0.15	$	&	$	5	\pm	1	$	&	$	17.9	\pm	0.2	$	\\
95108-01-10-00	&	$	4.86	\pm	0.04	$	&	$	0.47	\pm	0.14	$	&	$	4.1	\pm	0.8	$	&	$	18.1	\pm	0.3	$	\\
95108-01-11-00	&	$	4.64	\pm	0.03	$	&	$	0.42	\pm	0.10	$	&	$	4.9	\pm	0.8	$	&	$	18.9	\pm	0.2	$	\\
95108-01-12-00	&	$	4.78	\pm	0.03	$	&	$	0.57	\pm	0.10	$	&	$	10	\pm	2	$	&	$	17.8	\pm	0.2	$	\\
95108-01-13-00	&	$	6.13	\pm	0.06	$	&	$	0.79	\pm	0.24	$	&	$	2.8	\pm	0.6	$	&	$	15.5	\pm	0.2	$	\\
95108-01-14-00	&	$	5.01	\pm	0.05	$	&	$	0.56	\pm	0.16	$	&	$	4.2	\pm	1.0	$	&	$	18.3	\pm	0.3	$	\\
95108-01-15-00	&	$	5.10	\pm	0.03	$	&	$	0.60	\pm	0.11	$	&	$	3.9	\pm	0.4	$	&	$	17.7	\pm	0.2	$	\\
95108-01-16-00	&	$	6.02	\pm	0.04	$	&	$	0.88	\pm	0.16	$	&	$	6.6	\pm	0.9	$	&	$	15.2	\pm	0.1	$	\\
95108-01-17-00	&	$	7.01	\pm	0.09	$	&	$	0.98	\pm	0.32	$	&	$	3.3	\pm	0.8	$	&	$	12.9	\pm	0.1	$	\\
95108-01-18-00	&	$	7.5	\pm	0.3	$	&	$	0.78	\pm	1.28	$	&	$	1	\pm	1	$	&	$	11.7	\pm	0.2	$	\\
95108-01-18-01	&	$	7.3	\pm	0.2	$	&	$	1.01	\pm	0.63	$	&	$	2	\pm	1	$	&	$	12.4	\pm	0.2	$	\\
95108-01-19-00	&	$	7.2	\pm	0.2	$	&	$	0.69	\pm	0.55	$	&	$	1.0	\pm	0.5	$	&	$	13.1	\pm	0.3	$	\\
95108-01-20-00	&	$	6.6	\pm	0.1	$	&	$	0.79	\pm	0.30	$	&	$	1.7	\pm	0.5	$	&	$	14.6	\pm	0.3	$	\\
95108-01-21-00	&	$	7.2	\pm	0.2	$	&	$	1.18	\pm	0.74	$	&	$	1.3	\pm	0.7	$	&	$	11.0	\pm	0.1	$	\\
95108-01-22-00	&	$	6.34	\pm	0.04	$	&	$	0.63	\pm	0.11	$	&	$	3.1	\pm	0.4	$	&	$	14.2	\pm	0.1	$	\\
95108-01-23-00	&	$	5.33	\pm	0.04	$	&	$	0.57	\pm	0.12	$	&	$	5.6	\pm	0.8	$	&	$	17.4	\pm	0.2	$	\\
95108-01-24-00	&	$	5.99	\pm	0.04	$	&	$	0.69	\pm	0.11	$	&	$	4.1	\pm	0.5	$	&	$	15.9	\pm	0.1	$	\\
95108-01-25-00	&	$	5.81	\pm	0.04	$	&	$	0.60	\pm	0.13	$	&	$	4.1	\pm	0.6	$	&	$	15.5	\pm	0.2	$	\\
95108-01-26-00	&	$	6.67	\pm	0.06	$	&	$	0.81	\pm	0.19	$	&	$	2.6	\pm	0.6	$	&	$	14.1	\pm	0.1	$	\\
95108-01-27-00	&	$	3.90	\pm	0.49	$	&	$	0.49	\pm	0.03	$	&	$	2.8	\pm	0.1	$	&	$	6.5	\pm	0.2	$	\\
95108-01-28-00	&	$	4.09	\pm	0.05	$	&	$	0.60	\pm	0.13	$	&	$	2.1	\pm	0.3	$	&	$	8.4	\pm	0.2	$	\\
95108-01-30-00	&	$	6.01	\pm	0.06	$	&	$	0.56	\pm	0.17	$	&	$	1.3	\pm	0.3	$	&	$	15.5	\pm	0.3	$	\\
95118-01-01-00	&	$	6.85	\pm	0.08	$	&	$	0.81	\pm	0.23	$	&	$	1.6	\pm	0.4	$	&	$	13.3	\pm	0.1	$	\\
95118-01-01-01	&	$	3.85	\pm	0.04	$	&	$	0.53	\pm	0.11	$	&	$	1.7	\pm	0.3	$	&	$	6.9	\pm	0.4	$	\\
95118-01-02-00	&	$	3.47	\pm	0.09	$	&	$	1.21	\pm	0.28	$	&	$	2.0	\pm	0.3	$	&	$	5.1	\pm	0.3	$	\\
95118-01-06-00	&	$	6.77	\pm	0.85	$	&	$	5.12	\pm	2.87	$	&	$	1.5	\pm	0.5	$	&	$	6.0	\pm	0.5	$	\\
95118-01-07-00	&	$	3.30	\pm	0.04	$	&	$	0.48	\pm	0.12	$	&	$	1.5	\pm	0.3	$	&	$	8.4	\pm	0.5	$	\\
95118-01-09-00	&	$	2.39	\pm	0.23	$	&	$	1.37	\pm	0.96	$	&	$	1.4	\pm	0.9	$	&	$	7.2	\pm	0.7	$	\\
95118-01-10-00	&	$	3.55	\pm	0.04	$	&	$	0.37	\pm	0.09	$	&	$	2.9	\pm	0.5	$	&	$	9.6	\pm	0.3	$	\\
95118-01-11-00	&	$	1.74	\pm	0.39	$	&	$	1.07	\pm	1.39	$	&	$	0.8	\pm	0.7	$	&	$	6.6	\pm	1.4	$	\\
95118-01-12-00	&	$	2.02	\pm	0.12	$	&	$	0.65	\pm	0.43	$	&	$	1.4	\pm	0.7	$	&	$	7.3	\pm	0.6	$	\\
95118-01-16-01	&	$	5.97	\pm	0.09	$	&	$	0.69	\pm	0.37	$	&	$	1.5	\pm	0.6	$	&	$	14.8	\pm	0.4	$	\\
95118-01-17-00	&	$	4.78	\pm	0.06	$	&	$	0.56	\pm	0.18	$	&	$	2.2	\pm	0.5	$	&	$	18.9	\pm	0.4	$	\\
95118-01-17-01	&	$	3.36	\pm	0.05	$	&	$	0.47	\pm	0.12	$	&	$	3.2	\pm	0.6	$	&	$	23.8	\pm	0.4	$	\\
95118-01-18-00	&	$	2.58	\pm	0.08	$	&	$	0.28	\pm	0.15	$	&	$	1.4	\pm	0.5	$	&	$	24.4	\pm	1.1	$	\\
95118-01-19-00	&	$	2.19	\pm	0.03	$	&	$	0.43	\pm	0.08	$	&	$	4.2	\pm	0.6	$	&	$	27.0	\pm	0.4	$	\\
95118-01-20-00	&	$	2.04	\pm	0.04	$	&	$	0.33	\pm	0.10	$	&	$	3	\pm	1	$	&	$	26.3	\pm	0.5	$	\\
95118-01-21-00	&	$	1.63	\pm	0.04	$	&	$	0.52	\pm	0.16	$	&	$	5	\pm	1	$	&	$	27.2	\pm	0.5	$	\\

\end{longtable}																															
\end{center}
%\end{table*}
\twocolumn

%%_____________________PDS_TABLE____________________________%%

\label{lastpage}
\end{document}